\documentclass[12pt]{article}
\usepackage{graphicx, amsmath, amssymb, cite, setspace, color, relsize, bm, bbold, array, dsfont, xcolor, cancel,soul,placeins,datetime}
\usepackage{comment}

\usepackage[top=1 in, bottom=1 in, left=.9 in, right=.9 in]{geometry}

\newcommand{\captionfonts}{\footnotesize} 
\makeatletter
\long\def\@makecaption#1#2{%
  \vskip\abovecaptionskip
  \sbox\@tempboxa{{\captionfonts #1: #2}}%
  \ifdim \wd\@tempboxa >\hsize
    {\captionfonts #1: #2\par}
  \else
    \hbox to\hsize{\hfil\box\@tempboxa\hfil}%
  \fi
  \vskip\belowcaptionskip}
\makeatother

\def\lsim{ \lower .75ex \hbox{$\sim$} \llap{\raise .27ex
\hbox{$<$}} }
\def\gsim{ \lower .75ex \hbox{$\sim$} \llap{\raise .27ex
\hbox{$>$}} }

\let\oldsqrt\sqrt
\def\sqrt{\mathpalette\DHLhksqrt}
\def\DHLhksqrt#1#2{%
\setbox0=\hbox{$#1\oldsqrt{#2\,}$}\dimen0=\ht0
\advance\dimen0-0.2\ht0
\setbox2=\hbox{\vrule height\ht0 depth -\dimen0}%
{\box0\lower0.4pt\box2}}

\begin{document}

\title{The Channel Capacity of a Relativistic String}

\author{Adam~R.~Brown  \\
 \textit{\small{Google DeepMind, Mountain View, CA 94043, USA}} \\
 \textit{\small{Physics Department, Stanford University, Stanford, CA 94305, USA}} }
\date{}
\maketitle


\begin{abstract}
\noindent I explore the limitations on the capacity of a relativistic channel to transmit power and information that arise because of the finiteness of the transverse speed of light. As a model system, I consider a rope constructed from a fundamental string, for which relativistic invariance is built in. By wiggling one end of the string, both power and information may be transmitted to the other end.  I argue that even though an unbounded amount of power and information may be traveling down the string, there is a bound on how much may be transmitted. Further, I conjecture that the two kinds of channel capacity---power and information---interfere with each other, so that the only way to transmit the maximum amount of power is to send no information, and vice versa. 
\end{abstract}

\vspace{10cm} 

\thispagestyle{empty} 

\newpage

\maketitle

\section{Strings as channels}
\begin{figure}[htbp] 
   \centering
   \includegraphics[width=5in]{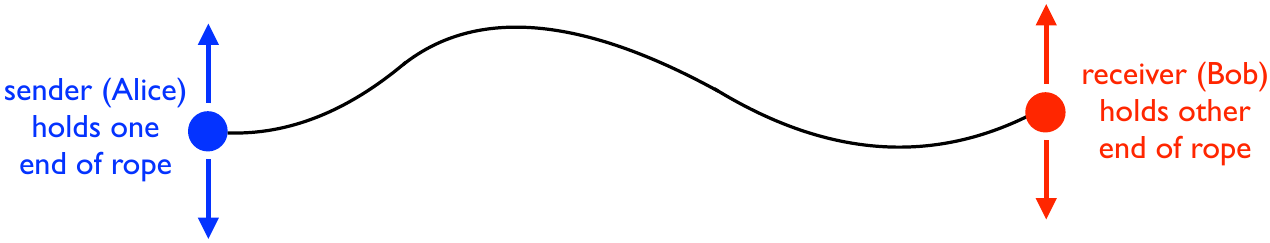} 
   \caption{Alice transmits power and information by wiggling the rope. The waves travel down the rope to Bob.}
   \label{fig:AliceAndBob}
\end{figure}

Alice and Bob can communicate only via a rope. Alice is holding one end of the rope and Bob is holding the other, as depicted in Fig.~\ref{fig:AliceAndBob}. It is clear that Alice can  send both power and information to Bob. But how much? 

The answer is going to depend on the properties of Alice, Bob, and the rope. If Alice is too weak to shake the rope, she may not be able to transmit much of anything. If the rope is too fragile, then even if Alice is strong she must be gentle lest she destroy the channel. Even if both Alice \emph{and} the rope are strong, Bob must also cooperate so as to receive (and not reflect) what Alice sends him. 
Since the focus  of our study will not be the contingent details of Alice, Bob, or the rope, we are going to ask only about those limitations on their capabilities that are imposed by the laws of physics. 

The specific law of physics that will be their principal antagonist is the law that says that nothing can travel faster than light. This limitation will manifest itself in three ways:
\begin{enumerate}
\item The speed of light limits the `longitudinal' speed of perturbations.

Signals take a time $\Delta x/c$ to reach Bob. This is the fact that takes the spontaneity out of conversations with Martians. This is the most obvious and least interesting limitation imposed by the finiteness of the speed of light, a limitation that introduces a time delay but does not in-of-itself limit the channel capacity, and a limitation that will not be the focus of this study.  
\item The speed of light limits the `transverse' speed of perturbations.

If Alice and Bob are allowed to move in any direction, then Alice can just walk over to Bob and give him all her power and information. We will forbid this. Instead, we will insist that Alice and Bob can only move in directions perpendicular (transverse) to their separation. 
The speed of light limits transverse motion. Alice will be limited by her ability to generate waves---and Bob will be limited in his ability to absorb them---by the fact that the endpoints of the rope cannot travel faster than the speed of light. 

\item The speed of light limits the strength of the rope. 

The null energy condition limits the strength of any rope in terms of its mass density and the speed of light: it requires that the tension be less than or equal to its mass per unit length times the speed of light squared, $\mu \hspace{1pt} c^2$.  One way to think about this limit is that the tension measures how much energy you need in order to stretch the rope a little, $d E = T dx$, and this cannot be larger than the cost to manufacture an extra segment of rope to place on the end, $d E = \mu c^2 dx$; another is that a rope with $T = (-w) \mu c^2$ and $(-w) > 1$ would have a speed of sound faster than the speed of light, $c_s^2 = \frac{\partial T}{\partial \mu} = (-w)c^2$. Either way, the finiteness of the speed of light  imposes a fundamental limit\footnote{In S.I.~units, $T/\mu \leq c^2 = 9 \times 10^{16}N/(kg/m)$, about a hundred million times stronger than carbon nanotubes.} on the tensile strength \cite{Brown:2012un}. In order to be limited only by the laws of physics, we will consider a rope that saturates the null energy condition, and so is as strong as any rope can be. A `rope' with \vspace{-1mm}
\begin{equation}
\textrm{tension} = \mu \hspace{1pt} c^2 \label{eq:tensionismassperunitlength} 
\end{equation}
and no lateral pressure is known as a `string'. Strings have been well-studied in the high-energy physics literature \cite{GreenSchwarzWitten}.  The string has relativistic invariance---both longitudinal and transverse---built in. 
\end{enumerate}

\noindent For simplicity, we take Alice and Bob to live in a fixed 2+1-dimensional Minkowski space, 
\vspace{-1mm}
\begin{equation}
ds^2 = -c^2 dt^2 + dx^2 + dy^2.  \label{eq:Mink2metric} \vspace{-1mm}
\end{equation}
Alice and Bob are subject to the restrictions of causality, unitarity, and  the limitation that they are not allowed to move in $x$. Alice is confined to $x = 0$. She may move to any value of $y(x=0)$, but may not exceed the speed of light, $|\dot{y}(x=0) |\leq c$. Bob is confined to $x=L$. He may move to any value of $y(x=L)$, but may not exceed the speed of light, $|\dot{y}(x=L) |\leq c$. 

Neither power nor information-rate are Lorentz invariants. To define them, we will pick a Lorentz frame. The Lorentz frame we will pick is the one that is at rest relative to the coordinates of Eq.~\ref{eq:Mink2metric}, so that $\dot{x} = \dot{y} = 0$. In this frame, we will ask three questions: 
\begin{enumerate}
\item How much power can Alice send to Bob?
\item How much information can Alice send to Bob?

(We will start with how many classical bits per unit time, but since this is a quantum channel we may also be interested in how many qubits per unit time.)

\item How much power and information Alice can send to Bob?

 (Can we max-out both rates, or is there are tradeoff? If so, what is the Pareto frontier?)
 \end{enumerate}

\noindent We will investigate each of these questions in turn. For the first question I will be able to provide a definitive answer: I will prove the upperbound on power transmission for a classical string and describe a strategy that saturates it. However, that will be the last thing I will be able to prove. For the second and third questions I will be able to describe neither the optimal encoding strategy nor its rate. Instead, as an invitation to further study, I will make a sequence of conjectures.

\section{How much \emph{power} can be sent down a string?} \label{sec:powerdownstring}

Alice imparts kinetic energy to the string by wiggling her end of the rope, the waves travel down the string, and then Bob extracts kinetic energy by absorbing the incoming wave. How much power can they transfer?

Let's examine the rope. The action of a rope that obeys Eq.~\ref{eq:tensionismassperunitlength} (and has no additional dependence on the extrinsic curvature) is equal to the spacetime area it sweeps out, 
\begin{equation}
S =  - \mu c^2 \int_\textrm{string} \hspace{-2mm}  \sqrt{-h} = - \mu c^2 \int dt dx \sqrt{ 1 - c^{-2} \dot{y}^2 +({y}')^2 } . \label{eq:NambuGoto}
\end{equation}
In this section we will take the string to be completely classical; we will discuss the effects of quantum mechanics in later sections. It follows from the action that the classical equations of motion are $y'' ( c^2 -\dot{y}^2 ) -  \ddot{y} ( 1 + y'^2 ) + 2 \dot{y} y' \dot{y}' = 0$. Solutions to these equations of motion include arbitrary right-moving waves $y(x,t) = y_r(ct-x)$, arbitrary left-moving waves $y(t,x) = y_l(ct+x)$, but not in general the sum of right-moving and left-moving waves. The energy of the string is 
\begin{equation}
E = \mu c^2 \int_0^L dx \frac{1 + (y')^2 }{\sqrt{ 1 - c^{-2} \dot{y}^2 + (y')^2} }  . 
\end{equation}
For a static string $\dot{y} = 0$, the energy is $\mu c^2$ times the length. For a string carrying only right-moving waves, the energy is 
\begin{equation}
E \Bigl|_{y(x,t) = y(ct-x)} =   \mu c^2 \int_0^{L} dx (1 + (y')^2) . \label{eq:energyinpurerightmoving}
\end{equation}

A nonrelativistic rope may carry kinetic energy both in the `transverse' direction (perpendicular to the rope) and in the `longitudinal' direction (along the rope). Both may in principle be used to transmit power. By contrast ropes that have relativistic tension $T = \mu c^2$ carry only transverse kinetic energy. Just as a photon has no longitudinal polarization, and just as an electric field line has no longitudinal rest frame, so a string has a longitudinal-boost-invariant stress tensor  and cannot carry longitudinal kinetic energy. This means the only way to transmit power with strings will be to wiggle them in the transverse direction.

It is clear that if Alice doesn't move her hand at all, then she will not send any power. Symmetrically, if Bob doesn't move his hand at all, then he will not absorb any power---he will be imposing a Dirichlet boundary condition $\dot{y}_\textrm{Bob} (t) \equiv \dot{y}(L,t)= 0$ and will therefore reflect back all incoming waves. Alice and Bob's ability to transmit and absorb power is determined by their ability to move their end of the string, and that is limited by the speed of light.

Here is an example of a power-transmission strategy.  Alice pulls her end of the string upwards at constant velocity, ${y}_\textrm{Alice}(t) \equiv {y}(0,t)  = \beta c  t$, tugging against the tension. Bob in turn is dragged along by Alice, following her upwards at the same velocity, ${y}_\textrm{Bob} = \beta(ct - L)$.  In order not to be accelerated by the string, and instead maintain constant velocity, Bob applies regenerative braking and extracts power. If Bob allows himself to be pulled along like this, then he extracts all the energy that Alice fed in, with no reflection. This means the shape of the string between them remains a pure right-moving wave $y(x,t) = \beta( ct  - x)$. The total energy in the rope is given by Eq.~\ref{eq:energyinpurerightmoving} as $E = \mu c^2 L ( 1 + \beta^2)$. 

Let's calculate the power transfer. To aid our accounting, let's imagine that at a certain moment Alice stops moving. At first Bob won't notice. As far as Bob is concerned, until a light-crossing time $L/c$ has passed, everything will proceed as it would have done had Alice not stopped. During this period he will absorb all the extractable energy in the rope. At the exact end of the period he will have exactly caught up with Alice, $y_\textrm{Bob} = y_\textrm{Alice}$,  leaving just a horizontal static rope of energy $\mu c^2 L$. He has thus absorbed energy $\mu c^2 L \beta^2$ in time $L/c$, so the power is $\mu c^3 \beta^2$. However, this is exactly the same amount of energy Bob would have absorbed had Alice \emph{not} stopped. So the constant-velocity strategy must transfer a power 
\begin{equation}
\mathcal{P} = \mu c^3 \beta^2 \ . 
\end{equation}

As $\beta \rightarrow 1$, this strategy transfers $\mathcal{P} = \mu c^3$. In this limit, Alice moves with a velocity $c$ against a force $\mu  \hspace{1pt} c^2$ and  exerts a power equal to the force times the velocity. Indeed, not only is this the maximum power Alice can transfer to Bob with this strategy, this is the maximum power she can transfer with \emph{any} strategy, as we now argue. No matter how Alice moves her hand, there is clearly no advantage to be had in Bob reflecting any power if he can avoid it. This means he should implement completely absorbing boundary conditions, which he can do by moving his end of the string in lockstep with Alice except delayed by a light-crossing time, 
\begin{equation}
\textrm{absorbing boundary condition:} \ \ \ \ \ \ \  \ \ {y}_\textrm{Bob}(t) = {y}_\textrm{Alice}(t-L/c) \ . \ \ \ \ \ \ \ \ \ \  \hspace{1cm} \label{eq:optimalpowerstrategy}
\end{equation}
If Bob moves his hand in this way then there is no reflection and so no left-moving waves. This means the energy in the string is given by Eq.~\ref{eq:energyinpurerightmoving},  which implies Bob extracts a power $\mathcal{P}(t) = \mu c \hspace{1pt} \dot{y}_\textrm{Bob}^2(t)$. Using $\dot{y}_\textrm{Bob}^2(t) \leq c^2$ gives a bound on the total power transfer by a classical string of mass-per-unit-length $\mu$ of 
\begin{equation}
\boxed{\mathcal{P} \leq \mu c^3 } \ . \label{eq:powerlimitconjecture}
\end{equation}
To saturate this bound, it is required that Alice moves her end always at the speed of light, but she is permitted to change direction.  Classically, Alice can move her end of the string either up or down, and can switch between up and down as often as she likes, so long as $|\dot{y}_\textrm{Alice}| = c$. Then if Bob moves his end of the string in accordance with Eq.~\ref{eq:optimalpowerstrategy} he will absorb all the power Alice sends. With a completely absorbing boundary condition, there is no reflection,  no left-moving modes, and the shape of the string is ${y}(x,t) = {y}(0,t-x/c)$. This shape will bear the zig-zag imprint of any changes of direction Alice might have made, as in Fig.~\ref{fig:triangularwaves}.

\begin{figure}[htbp] 
   \centering
   \includegraphics[width=5in]{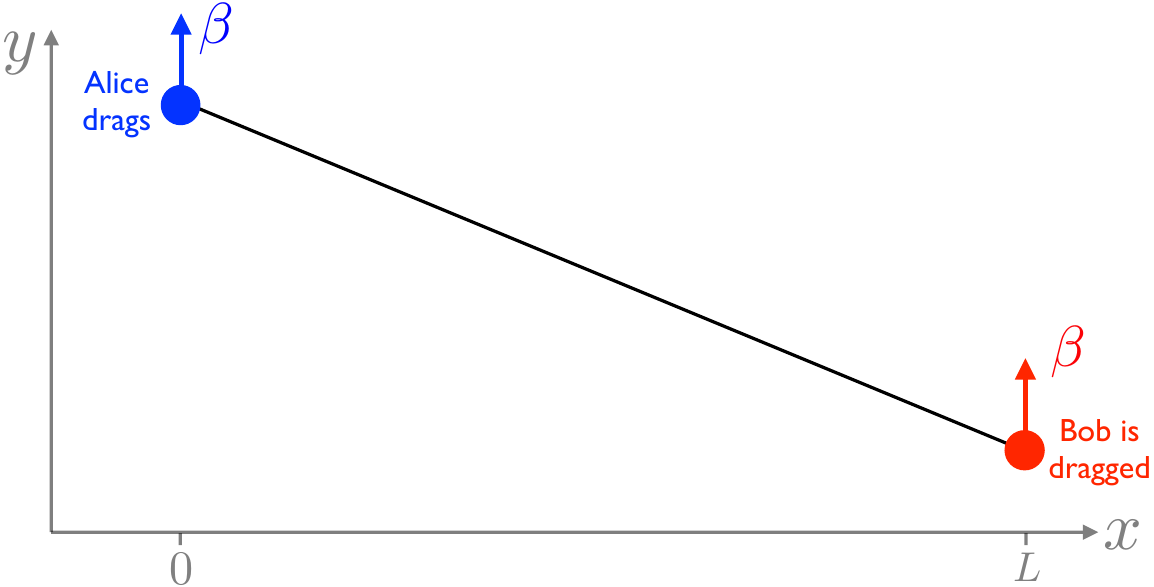} 
   \caption{A example of a power-transfer strategy. Alice drags her end of the string upwards, $y_\textrm{Alice} = \beta c  t$. Bob is dragged along behind her $y_\textrm{Bob} = \beta(c t - L)$. The shape of the rope between them is $y = \beta(c t -x)$. The power transferred is $\mu \beta^2 c^3$, and in the limit $\beta \rightarrow 1$ this saturates the bound Eq.~\ref{eq:powerlimitconjecture}. As $\beta$ approaches 1, the string lies at a $45^\circ$ angle in our frame, but that is because of the length contraction of $y$: in \emph{their} Lorentz frame the string looks almost vertical.}
   \label{fig:dragrace}
\end{figure}

Finally, let us remark that even though  only a limited amount of power can be \emph{fed in} by Alice or \emph{extracted by} Bob---due to the limitation that they cannot move their respective ends of the string faster than the speed of light---there is no limit on the transverse power that can be \emph{traveling down} the string. If a right-moving mode has power $\mathcal{P}$ in one frame, then in another frame longitudinally boosted with $x$-velocity $-\beta c$ the power will be blueshifted, 
\begin{equation}
\mathcal{P}' = \frac{1 + \beta}{1 - \beta} \mathcal{P} , 
\end{equation}
which can be made arbitrarily large with sufficient boosting. For example, if Alice is boosted in the $+x$ direction, then she can create a right-moving mode that has shape $y = \sqrt{\frac{1 + \beta}{1 - \beta}} (ct - x)$, which in the lab frame has $\mathcal{P}' = \frac{1 + \beta}{1 - \beta} \mu c^3 > \mu c^3$ and $dy/dt = \sqrt{\frac{1 + \beta}{1 - \beta}} c > c$. This does not violate causality, since Bob cannot use this incoming string to communicate faster than light, nor can he extract more power from the string than Eq.~\ref{eq:powerlimitconjecture}. Instead, confined to move with $|\dot{y}_\textrm{Bob} |\leq c$, he will end up reflecting the excess power back towards Alice (and in the process  lengthening the string).
In summary, there is no limit on the amount of power that can be traveling down a classical relativistic string, but Alice can't {send} more power, and Bob can't {extract} more power (in the $\dot{x} = \dot{y} = 0$ frame), than $\mu c^3$.

\section{How much \emph{information} can be sent down a string?} \label{sec:infodownstring}

Now suppose that 
instead of transmitting power, Alice wants to send Bob a message. She wiggles her end of the string so as to encode her message, and then Bob observes those wiggles arriving at his end and decodes the message.  For now let's take the message to be classical and written as a binary sequence 
of 0s and 1s.

\subsection{Classical string} \label{subsec:classicalstringcoding}
The channel capacity of a classical string is infinite. 
The location of a classical string as it passes a given value of $x$ is an analog variable---a function of time $y_x(t)$. Specifying an infinite-precision function, even one bounded to have $|\dot{y} |\leq c$, requires an infinite number of bits. Indeed, even specifying the value of $y_x(t)$ at a single value of $x$ \& $t$---i.e.~specifying a real number---requires an infinite number of bits. 
Since by assumption there is no noise, Alice can input an arbitrarily large amount of information into an arbitrarily short section of string, and Bob can reliably extract it. 

We can see the same thing starting with the Shannon-Hartley theorem \cite{Shannon:1948dpw}. The Shannon-Hartley theorem says that the information-transmission capacity $\mathcal{I}$  of a classical analog variable with bandwidth $B$,  average signal power $S$, and Gaussian noise power $N$, is  
\begin{equation}
\mathcal{I} = B \log \left( 1 + \frac{S}{N} \right) \ . \label{eq:ShannonHartley}
\end{equation}
The Shannon-Hartley theorem applies to a nonrelativistic string, and we can ensure the string is nonrelativistic almost everywhere by taking a modest range of frequencies (modest $B$) and inputting only a modest power (modest $S$). Because for a classical string there is no noise, $N=0$, this formula then tells us that Alice can communicate unboundedly large amounts of information to Bob. 

\begin{figure}[htbp] 
   \centering
   \includegraphics[width=5in]{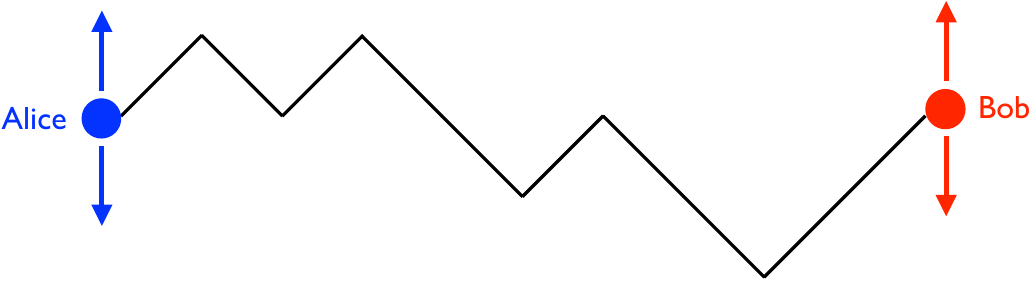} 
   \caption{`Triangular' coding scheme. Alice moves her end of the string always at the speed of light. To encode $1$, she pulls the string up at the speed of light; to encode 0, she pulls the string down at the speed of light. After a time $\delta$ she moves on to the next bit. The bits are thus spaced a horizontal distance $\Delta x  = c \delta $ apart, and each has vertical range $\Delta y |_\textrm{signal} = c \delta$.}
   \label{fig:triangularwaves}
\end{figure}

Let's see this for an explicit example of a coding scheme. Consider the `triangular' code shown in Fig.~\ref{fig:triangularwaves}. For each time period, when Alice wishes to transmit `1' she raises her end of the string at the speed of light $\dot{y}_\textrm{Alice} =+c$; when Alice wishes to transmit  `0' she lowers her end at the speed of light $\dot{y}_\textrm{Alice} = -c$. After a time $\delta$ Alice moves on to the next bit. Bob observes the shape of the wave as it arrives at his end and infers the message. Classically there is no  noise, so every bit gets through unscathed and the bit-rate is just the inverse of the time between bits, $\mathcal{I} =\delta^{-1}$. By reducing $\delta$, Alice and Bob can communicate bits at an unbounded rate. The channel capacity of a classical string is infinite.

\subsection{Quantum string} 

Now let's consider the impact of quantum mechanics. We will see that quantum fluctuations in the string can make it hard for Bob to read Alice's message. We will argue that this quantum `noise' makes the channel capacity  finite.  

Let's try to estimate that channel capacity. In the limit $L \rightarrow \infty$,  the only quantity left with the dimensions of bandwidth is the inverse `string time', $c/\ell_s \equiv \sqrt{{ \mu c^3}/{\hbar}}$. (The `string length' $\ell_s \equiv \sqrt{\hbar/\mu c}$ is the length-scale shorter than which the string is highly quantum, and should not be confused with the length of the string $L$; in this paper we will always assume $L \gg \ell_s$.) This means that, assuming the channel capacity is indeed neither zero nor infinity as $L \rightarrow \infty$, dimensional analysis suggests we should expect the channel capacity to be a multiple of $c/\ell_s$.

To support that hypothesis, let's re-examine the `triangular' coding scheme from Fig.~\ref{fig:triangularwaves}. We will argue that, even including the effects of quantum mechanics, it still works when the time between bits $\delta$ is a large multiple of the string time, giving a bit rate that is a small multiple of the inverse string time. Conversely, we will give a heuristic description of the various things that can go wrong if Alice and Bob try to use the triangular code with $\delta$ less than $\ell_s/c$.


The first thing that goes wrong 
 is that Bob simply cannot reliably tell which way Alice moved her hand. Quantum fluctuations in the transverse location of the string obscure its exact location. This means that if Bob measures  his end of the string to have moved up by a tiny amount,  he cannot be sure whether that was caused by Alice deliberately moving her end of the string up (in order to send the bit `1'), or just by random quantum jitter. The location of the string $y(x,t)$ is now not a classical function but a quantum field, and so as Alice's wave approaches Bob it is now in a superposition over many different values of $y$ (and typically entangled with the value of $y(x,t)$ at other values of $x$). 
It is a bad idea for Bob to try to measure the value of $y(x)$ at any particular $x$, since eigenstates of $\hat{y}(x)$ have UV-divergent energy. Instead Bob must smooth in both $y$ and $x$, doing a finite-precision measurement of the \emph{average} value of $y(x)$ smoothed over some interval.
Let's be quantitative. Consider a segment of the string of length $c \delta$.  What is the spread in the average transverse location? 

First consider the quantum fluctuations on a static horizontal string. So long as the perturbations have $ \dot{y}\ll c$ \& $y' \ll 1$, the energy of the segment of string is given by 
\begin{equation}
E =  \mu c^2  \int_0^{c \delta} dx   \frac{1 + (y')^2 }{\sqrt{ 1 - c^{-2} \dot{y}^2 + (y')^2} }  =   \mu c^2  \int_0^{c \delta} dx   \left( 1 + \frac{1}{2} c^{-2} \dot{y}^2 + \frac{1}{2} (y')^2 +  \ldots \right)  \ . \label{eq:energyofstringnonrel}
\end{equation}
The  modes that do not change the endpoints of the string are $y = \sum_n A_n \sin \frac{n \pi x}{L}$. The mode that contributes the most to the dispersion of the $x$-{averaged} $y$-location of the string is the $n=1$ mode. Inserting this mode into Eq.~\ref{eq:energyofstringnonrel} gives $E = \frac{1}{4} \mu c \delta  ( \dot{A}_1^2 + \pi^2 \delta^{-2} A_1^2 + \ldots )$. This is a simple harmonic oscillator with  $m=\frac{1}{2}\mu c \delta $ and $\omega = \pi /\delta$ so the groundstate has 
\begin{equation}
\langle A_1^2 \rangle \sim \frac{\hbar}{\mu c} \equiv \ell_s^2 \ \ \ \ \ \ \& \ \ \ \ \langle \dot{A}_1^2 \rangle \sim \frac{\hbar }{\mu c \delta^2} \equiv \frac{\ell_s^2 }{\delta^2} \ \ \ \ \ \ \& \ \ \ \  \langle E \rangle \sim \frac{\hbar }{\delta} \ \ \ . \label{eq:quantumspreadingroundstate}
\end{equation}
Even in its quantum mechanical groundstate, the string thus has a quantum `width' from quantum fluctuations of its fundamental mode, 
\begin{equation}
\ \ \ \ \ \ \ \ \ \  \Delta y \Bigl|_\textrm{noise}  \  \  \sim \  \ \ell_s \, \ \equiv \  \sqrt{ {\hbar}/{\mu c}}  \label{eq:quantumwidth}  \ , 
\end{equation}
independent of $\delta$.  This result is reliable so  long as $c \delta$ is much longer than $\ell_s$, because then the perturbations are self-consistently nonrelativistic $\dot{y}\ll c$ \& $y' \ll 1$. 
We can repeat this analysis when the unperturbed string is not static but moving as a right-going wave, for example $y(x,t) = \beta(ct - x)$. In the rest frame of the string, the transverse perturbations must still have typical size $\Delta x_{\perp} \sim \ell_s$. We wish to translate from transverse perturbations in the boosted frame to $\Delta y$ perturbations in the lab frame. We get one factor of $\gamma$ from the fact the string is almost vertical in the boosted frame, $\Delta \bar{y} = \gamma \Delta x_{\perp}$, but then lose one factor of $\gamma$ from the length contraction when we boost back to the lab frame $\Delta y = \gamma^{-1} \Delta \bar{y}$.  Altogether, we find that the characteristic smearing in the vertical direction is about the same as it was for the static horizontal string, $\Delta y \sim \ell_s$. 

With this in mind, let's revisit the triangular coding scheme of Sec.~\ref{subsec:classicalstringcoding}. In a given time period $\delta$, the `signal' is the change in the $y$-location of the string that Alice can create through her deliberate actions, which given that she is limited by the speed of light is 
\begin{equation}
\Delta y \bigl|_\textrm{signal} \sim c \delta \ . 
\end{equation}
This signal has to compete with the quantum noise. The groundstate contribution to the quantum noise of the fundamental mode is given by Eq.~\ref{eq:quantumwidth}. For a coherent state, therefore, the signal can only be larger than the noise if
\begin{equation}
\textrm{signal } \gg \textrm{ noise \ \ \  only if } \ \ \ \delta \gg \ell_s / c \ . 
\end{equation}
There are other contributions to the noise from the other odd modes, but these are suppressed relative to the $n=1$ mode both by having larger $w \sim n/\delta$ and so smaller amplitude, and also by the fact that at fixed amplitude they contribute less to the average displacement of $y$. This means, that $\delta = \ell_s/100c$ will result in the signal being almost completely washed out, but $\delta = 100 \ell_s/c$ should suffice so that the great majority of the bits in the message make it through. (The remaining errors can be dealt with in the standard way using the noisy channel coding theorem \cite{Shannon:1948dpw}.) 
Since the bit-rate is $\mathcal{I} = \delta^{-1}$, this implies an upperbound on how much information can be sent down the string. This bound, which I conjecture holds for any communication scheme Alice and Bob may adopt not just the triangular code, is 
\begin{equation}
\boxed{\mathcal{I} \ \lsim \   \alpha \, c \,  \ell_s^{-1} \equiv \alpha \, \sqrt{\frac{ \mu c^3}{\hbar}}} , \label{eq:channelcapacity}
\end{equation}
where $\alpha$ is some O(1) constant that I have not been able to calculate. Looking at the right-hand side of this expression, we see that the bound becomes trivial {either} when $c \rightarrow \infty$ {or} when $\hbar \rightarrow 0$: for there to be an upperbound on the channel capacity of the string requires \emph{both} quantum mechanics (to stop the noise being zero) \emph{and} the finiteness of the speed of light (to stop the signal being infinite).

Let's see that a similar estimate holds if instead of doing a finite-resolution measurement in the position basis, Bob does a finite-resolution measurement in the momentum basis. It follows from the action Eq.~\ref{eq:NambuGoto} that the momentum of a segment of string is 
\begin{equation}
p_y = \mu \int_0^{c \delta} dx   \frac{  \dot{y}}{\sqrt{ 1 -  c^{-2} \dot{y}^2 +({y}')^2 } } \ . 
\end{equation}
For a triangular wave $y(x,t) = \pm \beta (ct - x)$ the signal strength is thus
\begin{equation}
\Delta p_y  \Bigl|_\textrm{signal} \sim \mu c^2 \beta\delta \ . 
\end{equation}
This is maximized by $\beta = 1$. 
This must compete with the momentum's quantum spread. Since the harmonic oscillator groundstate has $\Delta p_y \Delta y \sim \hbar$, we can use the $\Delta y \sim \ell_s$ result we computed above to find the quantum spread of momentum in the groundstate from the fundamental mode to be  
\begin{equation}
\Delta p_y \Bigl|_\textrm{noise} = \frac{\hbar}{\ell_s} \ . 
\end{equation}
Requiring that the signal be greater than the noise gives $\mu c^2 \delta  \gg \frac{\hbar}{\ell_s}$ and therefore recovers $c \delta  \gg \ell_s$ and Eq.~\ref{eq:channelcapacity}.

We have given two motivations for Eq.~\ref{eq:channelcapacity}, one from the perspective of $\Delta y$, and the other from the perspective of $\Delta p_y$. Let's give four more. All of these are heuristic, and take results derived for nonrelativistic (and sometimes classical) transverse perturbations and adapt them for relativity by crudely imposing a speed limit $c$. Nevertheless together they tell a coherent story. First, consider the fact that in order to distinguish one bit from the next, we must be able to tell when the signal arrived with time-resolution better than $\delta$. The (nonrelativistic) result in Ref.~\cite{Aharonov:1997md} tells us that measuring the timing with resolution $\delta$ needs energy $\hbar/\delta$. Requiring that this be no greater than the incoming energy in the  string segment, which Eq.~\ref{eq:powerlimitconjecture} bounds as  $\mu c^3 \delta$, implies $\delta > \ell_s/c$ and thus gives Eq.~\ref{eq:channelcapacity}.  Second, and now moving beyond the triangular coding scheme, we could start with the (nonrelativistic) theorem in Ref.~\cite{Caves:1994zz} that for a bosonic channel $\mathcal{I} < \sqrt{ \mathcal{P}/\hbar}$, and plug in the power bound from Eq.~\ref{eq:powerlimitconjecture} to recover Eq.~\ref{eq:channelcapacity}. Third, we could start with the Shannon-Hartley theorem, Eq.~\ref{eq:ShannonHartley}. The Shannon-Hartley theorem applies to a nonrelativistic (in the transverse direction) classical channel with Gaussian noise, but we can heuristically model the noise from the quantum spread in the groundstate, described in Eq.~\ref{eq:quantumspreadingroundstate}, as being approximately Gaussian, and then use the bound on the signal power from Eq.~\ref{eq:powerlimitconjecture}. This will tell us that we can arrange for the signal to be greater than the noise for frequencies smaller than the inverse string time, giving a bandwidth of about $c/\ell_s$ and again  recovering Eq.~\ref{eq:channelcapacity}. As a final way to derive Eq.~\ref{eq:channelcapacity}, we could apply the heuristic principle that the energy in the quantum excitations of the string should be no greater than the energy in the classical motion. In our analysis of the triangular coding strategy, we implicitly assumed that Alice prepares the string in a coherent state: each segment of string, when boosted to its rest frame, is in its quantum mechanical groundstate. Alice might instead be tempted to prepare the string in a squeezed state. The advantage of doing so would be that the dispersion $\Delta y$ can be smaller than the `standard quantum limit'---smaller than Eq.~\ref{eq:quantumwidth}. This would make it easier for the signal to stand out amidst the noise. The disadvantage of squeezed states (in addition to being difficult to prepare, and also forcing us to be extremely careful with timing because they oscillate between squeezing ($\Delta y \ll \ell_s$) and anti-squeezing ($\Delta y \gg \ell_s$) on a timescale of $\omega^{-1} \sim \delta$) is that they are highly energetic, and that this energy is highly `quantum', meaning there is a large dispersion in the energy. We can estimate the energy dispersion as follows. A squeezed $\Delta y$ means an anti-squeezed $\Delta p_y \sim \hbar / \Delta y$, which for a nonrelativistic perturbation on a segment of string of length $c \delta$ means  $\Delta E \sim  \Delta p_y^2/m \sim  \hbar^2  / \mu c \delta \Delta y^2  = (\hbar/\delta )(\ell_s^2/\Delta y^2)$. Thus the energy in the quantum excitation  above its groundstate of the segment of string is $\Delta E  \geq \mu c^3 \delta (\frac{\ell_s}{c \delta})^4$. If we demand that this is less than the classical energy in the string $\mu c^3 \delta$ then squeezing is counterproductive and we are back to demanding $c \delta > \ell_s$. If we allow the quantum excitation energy to be greater than the classical energy then we lose control of the calculation and Alice and Bob lose control of the string. Bob needs to be extremely careful not to reflect any of that excess energy back towards Alice, lest it collide with the incoming waves and interfere with her later messages. (We will explore more about the dangers of Bob reflecting energy in the next section.) 

In summary, though I have not proved it, it seems all attempts to transmit information down a string with a bit rate larger than about $c/\ell_s$ are defeated by quantum noise.

\section{Power and Information:  exclusion conjecture}
\label{sec:exclusionconjecture}

The string can transmit only a limited amount of power (Eq.~\ref{eq:powerlimitconjecture}) from Alice to Bob. The string can, I have claimed, transmit only a limited amount of information per unit time (Eq.~\ref{eq:channelcapacity}) from Alice to Bob. Let me now conjecture an `exclusion principle' that says we cannot use both of these capacities at once:
\begin{equation}
\textrm{conjecture: \ \ Alice and Bob cannot simultaneously saturate Eq.~\ref{eq:powerlimitconjecture} and Eq.~\ref{eq:channelcapacity}.} 
\end{equation}
There is no strategy that simultaneously sends maximum power and maximum information. 

Here is a simple intuition for why this should be so. If Bob wants to receive power he must move his hand: if he just keeps his hand fixed, $\dot{y} = 0$, he won't absorb, he'll reflect. As we have discussed, to implement perfectly absorbing boundary conditions at his end of the rope, Bob should mimic Alice, moving his hand when he receives the wave in exactly the same way that Alice moved her hand when she sent the wave, $y_\textrm{Bob}(t) = y_\textrm{Alice}(t-L/c)$, see Eq.~\ref{eq:optimalpowerstrategy}. But to be able to implement this boundary condition, Bob must know which way the string is moving before the string reaches him: he must know the incoming $y'(t,L)$. On the other hand, he cannot receive new information from a variable if he already knows the value of the variable. Alice and Bob thus appear to be faced with the following dilemma,
\begin{eqnarray}
\textrm{dilemma:} && \textrm{to receive power, Bob must know $y_\textrm{Alice}(t)$ in advance} \nonumber \\
&&  \textrm{to receive information, Bob must not know $y_\textrm{Alice}(t)$ in advance. } \ \ \ \ \ \ \ \  \nonumber 
\end{eqnarray}
If Bob knows the location of the string well enough to implement an absorbing boundary condition---if he knows where the string is going to move before it moves---the location of the string does not carry new information for Bob. Consequently Bob can receive maximum power or maximum information but not both.

If they are using the triangular coding scheme of Fig.~\ref{fig:dragrace} to send information, Bob cannot start to be confident about whether the string is moving up or down until a string time $\ell_s /c$ has passed. During this time he doesn't know which way to move his hand in order to enforce the absorbing boundary condition Eq.~\ref{eq:optimalpowerstrategy}. As a consequence, during this time he will reflect about $\mu c^2 \ell_s$ worth of incoming energy. This is bad for two reasons. It is bad directly, because he forgoes energy he could otherwise have absorbed. And it is bad indirectly, because the energy he missed out on doesn't just disappear, instead it is reflected back towards Alice where it can disrupt later incoming waves. Because the action, Eq.~\ref{eq:NambuGoto}, is non-quadratic, this makes the equation of motion non-linear, which means that right-moving and left-moving modes interact. The $\mu c^2 \ell_s = \hbar c/\ell_s$ worth of right-moving energy that Bob failed to absorb becomes left-moving energy. This left-moving energy delays future right-moving waves by about a string time $\ell_s/c$ \cite{Dubovsky:2012wk}, corrupts incoming messages by distorting and entangling with them, and must be carefully removed by Alice `cooling' her end of the string lest the left-moving waves get unintentionally reflected back towards Bob again and cause yet more disruption.

If my conjecture is correct, there is a tradeoff between transmitting power and transmitting information. There is a Pareto frontier that describes the maximum amount of information that can be transmitted per unit time at a given rate of average power transfer. Different encoding strategies will lie at different points on this Pareto frontier (or if they are suboptimal strategies, they will lie within the frontier, and will be dominated by other strategies that transfer both more power and more information). I do not know where the Pareto frontier lies, or even have a particularly firmly held conjecture. But there is one possibility that has a special feature, 
\begin{eqnarray}
\textrm{linear Pareto frontier:}  \hspace{2cm} \frac{1}{\mu c^3}  \mathcal{P}  + \frac{ \ell_s}{\alpha c} \mathcal{I} & \leq & 1 \ . \hspace{1cm} \label{eq:linearexclusionconjecture} 
 \end{eqnarray}
 A linear Pareto frontier would be neither convex nor concave. We know that the Pareto frontier cannot be concave, because for any two points on the frontier, we can always construct any strategy that lies on a straight line between them by alternating long periods of one strategy with long periods of the other strategy. If the Pareto frontier is linear, this would be  the best one could do. \\

I have conjectured that attempts to transmit power interfere with attempts to transmit information. This sounds like the opposite predicament to that usually afflicting telegraphers. For telegraphy, if you want to send more information, you typically have to increase the power you send down the line. The classical Shannon-Hartley theorem, Eq.~\ref{eq:ShannonHartley}, says that for a fixed line with fixed noise and fixed bandwidth, the only way to increase the amount of information is to increase the signal power $S$. Other, more quantum, bounds also imply that communication demands power \cite{Bekenstein:1981zz}.

These results are not inconsistent. The Shannon-Hartley theorem is a bound on how much power must be \emph{fed into} the telegraph wire. A similar kind of bound applies to a relativistic string---if you artificially limited the power that Alice can input (for example by bounding her $\beta$ at less than $1$), you'd find that the information she can transmit would be strictly less than Eq.~\ref{eq:channelcapacity}; conversely if you enhanced the amount of power that Alice can input (for example by boosting her in the $+x$ direction so that her signal is blueshifted in the lab frame) then she can exceed Eq.~\ref{eq:channelcapacity}. Indeed, one way to think about Eq.~\ref{eq:channelcapacity} is precisely that limited power fed into the string leads to limited information transferred, and there is a limit from the speed of light on the amount of power that Alice can feed in. The reason that none of this is in tension with our conjectured exclusion principle is that the telegraphy bounds are lowerbounds on how much energy must be fed in by Alice to achieve a certain information transfer, whereas our bound is an upperbound on how much energy Bob can extract. Because of the possibility---and indeed we have argued inevitability---of Bob reflecting much of the energy that is sent his way if he wants to receive information, these are not the same quantities. \\

\begin{figure}[htbp] 
   \centering
   \includegraphics[width=2.7in]{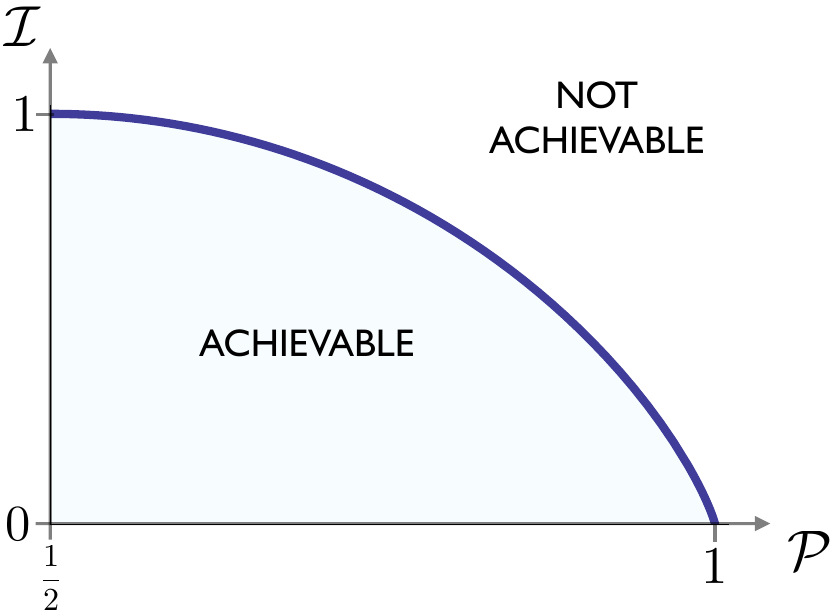}  
   \caption{The Pareto frontier for the discrete analogy, given by Eq.~\ref{eqParetoForOneBit}. This shows how much information $\mathcal{I}$ can be  communicated per iteration at a given rate of average power transfer $\mathcal{P}$. Everything below the line is achievable; everything above the line is unattainable.}
     \label{fig:ParetoOneBit}
\end{figure}

Let us make a discrete analogy. This discrete analogy will manifest some but not all of the characteristics of the case of a continuous relativistic string.  There is now no string. 
Instead Alice and Bob get to communicate exclusively via sequential access to a two-state system. The two states have an energy difference $\Delta E=1$. First Alice places the system in one of the two states. Then she hands it to Bob. Then Bob examines and interacts with it. Then he hands it back to Alice. Then the process repeats. 
To send maximum power, Alice should always hand Bob the excited state. That way Bob can always extract unit energy by draining it back to the groundstate. But if he knows in advance which state Alice will hand him, then observing that state conveys no information. There is a tradeoff. Let's calculate that tradeoff. If they choose a coding scheme for which a fraction $p$ of the states Alice hands Bob are excited, then the transmitted power is $\langle \mathcal{P} \rangle = p$, which is maximized by $p=1$. The rate of information transmission is $\langle \mathcal{I} \rangle = - p \log p - (1-p) \log (1-p)$, which is maximized by $p=\frac{1}{2}$. If Alice wants to send both power \emph{and} information, these two channels interfere: $p$ cannot be both $1$ and $\frac{1}{2}$. 
 The Pareto frontier, plotted in Fig.~\ref{fig:ParetoOneBit}, is 
\begin{equation}
\langle \mathcal{I} \rangle \   \leq \  -   \langle \mathcal{P} \rangle \log \langle \mathcal{P} \rangle - (1-\langle \mathcal{P} \rangle) \log (1-\langle \mathcal{P} \rangle) .  \label{eqParetoForOneBit}
\end{equation}

\section{Knowledge, Power, and the Superstring}
We have studied the maximum amount of power and information that may be transmitted from Alice to Bob using a relativistic string. In Sec.~\ref{sec:powerdownstring} we bounded the power. In Sec.~\ref{sec:infodownstring} we conjectured a bound on the information. In Sec.~\ref{sec:exclusionconjecture} we conjectured a joint bound---an exclusion principle on the simultaneous transmission of power and information.

 This paper sketches out a program, but leaves unfinished business. The first piece of unfinished business is proving a bound on the information transfer rate for a string. If my conjecture is correct, this will amount to finding the value of $\alpha$ in Eq.~\ref{eq:channelcapacity}, which will involve optimizing the mutual information over all possible communication strategies that Alice and Bob could employ. The second piece of unfinished business is deriving the shape of the Pareto frontier. I have not been able to derive either of these in this paper. 
 
 The first step in approaching these questions should be to make a quantitative model of the internal Hilbert spaces of Alice and Bob and the ways in which unitarity and causality limit how they can couple those Hilbert spaces to their respective ends of the string. All the arguments in this paper have been semiclassical (or classical), and an explicitly quantum description would confirm whether the semiclassical heuristics  are reliable.  For example, when we say that Alice are Bob are `holding' their ends of the rope, it is clear what this means in the classical context, but not clear how to translate that classical picture into a fully quantum description. Exactly what kinds of couplings are permitted? How much are  Alice and Bob able to spread themselves in the $y$-direction?
 
Let me give a couple of concrete examples of issues that an explicit quantum model of Alice and Bob would address. First, recall that Sec.~\ref{sec:powerdownstring} shows that classically there is a maximum power that Alice can input into the string; on the other hand, quantum mechanically, if Alice can  rapidly measure the location of a short string segment with extremely high precision, then she can place it in a squeezed state that has extremely high energy, corresponding to a huge power input. An explicit model of Alice's coupling to the string would tell us whether causality permits Alice to rapidly create a highly squeezed state, and an explicit model of Bob's coupling to the string would tell us whether, if so, Bob is able to extract that power. Similarly, imagine modifying the set-up of Fig.~\ref{fig:AliceAndBob} so that rather than ending at Bob, the string continues to Bob's right forever. If the string to Bob's right has the same $\mu$ as the string to Bob's left, there will be no impedance mismatch and no reflection. This is a perfectly absorbing boundary condition that does not require knowing the shape of the incoming wave in advance. Returning now to our set-up, Fig.~\ref{fig:AliceAndBob}, in which the string ends at Bob, an explicit quantum mechanical model  would tell us whether Bob is able to configure his internal Hilbert space and its couplings so as to simulate the existence of a semi-infinite string reaching out to his right. If he's able to do this, and do this in a way that does not rely on knowing in advance the shape of the incoming wave,  then it might imperil the conjectures of Sec.~\ref{sec:exclusionconjecture} and perhaps even Sec.~\ref{sec:infodownstring}.  This is the gravest threat that I see to the conjectures of this paper, and until it is addressed with an explicit quantum model the results of this paper can only be considered heuristic.  \\

We have considered the transmission of classical information. But a relativistic string may be put in quantum superposition, and can transmit quantum information.  It seems reasonable to conjecture that versions of Eqs.~\ref{eq:channelcapacity} and the results of Sec.~\ref{sec:exclusionconjecture} exist for quantum information, and that the coefficient $\alpha$ is not much smaller than it is for classical information. Relatedly, in this paper we have assumed that Alice and Bob share control of a string, but share no additional entanglement. One could investigate how giving them a supplemental entanglement resource would increase their ability to make use of the string. Similarly, we have only asked about one-way transmission (from Alice to Bob), but in principle the string could be used to transmit in both directions.

Now let's consider what happens if we are able to tease apart our string into $N$ lighter strings each of mass per unit length $\mu/N$. Examining Eq.~\ref{eq:powerlimitconjecture}, we see that this does not change the total transmitted power. By contrast,  Eq.~\ref{eq:channelcapacity} tells us that, if the strings are distinguishable, the amount of information that may be transmitted increases by a factor of $\sqrt{N}$.

Relatedly, we have assumed that there is only one transverse direction. We could increase that to $d-1$ transverse directions by placing the string in $d+1$-dimensional Minkowski. Additional transverse directions should not increase the capacity to send power, since the optimal power-transfer strategy will still be for Alice to choose a transverse direction---any transverse direction---and proceed in that direction at the speed of light. Additional transverse directions should, however, increase the information-transfer capacity of the string because Alice, when she generates her wave, can encode additional information in her choice of direction.  
 
A power-information exclusion principle has also been found in a forthcoming paper by Jinzhao Wang and Shunyu Yao in a seemingly different context, that of the quantum teleportation of power and information \cite{WangYao2024}. While there are a number of important differences between their set-up and mine, it would be interesting to investigate if there are deeper connections. 
 
Finally, let's address the theoretical consistency of our thought experiment. We described the behavior of a relativistic string in a fixed 2+1-dimensional Minkowski. This is perfectly consistent at the classical level, but quantum mechanics introduces two  dangers. The first danger is that the classical conformal symmetry on the 1+1-dimensional surface traced out by the string (the `worldsheet')   is anomalous---broken by quantum effects---which ultimately means that the string theory we have described is not UV complete and is at best a low-energy effective theory. We can deal with this either by just accepting that we have an effective theory and placing the cut off sufficiently high that it does not affect our results (recall that the string theory is free $g_s=0$ and the background is fixed), or by posing the question in 25+1 dimensions where the anomaly cancels and therefore conformal symmetry is unbroken. The second danger is that even when the conformal symmetry is not anomalous, the question we have asked is not well posed because the spectrum of the string has a tachyon. We could attempt to neutralize that tachyon by adding extrinsic curvature terms to the action, or again just ignoring it and treating our string theory as an effective theory, but the  most famous way to eliminate the tachyon is to add fermions to the worldsheet (which also necessitates moving to 9+1 dimensions). We could then investigate the adversarial relationship between knowledge, power, and the superstring. 
 
\section*{Acknowledgements} 
Thanks to Patrick Hayden, Henry Lin, Sandu Popescu, and Lenny Susskind for helpful feedback. The first draft of this paper was written at PrimoCosmo13 at KITP and I thank the hosts for their hospitality. I also thank Jinzhao Wang for encouraging me to finally disseminate these ideas and for sharing an advance copy of his simultaneously released Ref.~\cite{WangYao2024}.


\begin{thebibliography}{99}

\bibitem{Brown:2012un} 
  A.~R.~Brown,
  ``Tensile Strength and the Mining of Black Holes'',
  Phys.\ Rev.\ Lett.\  {\bf 111}, no. 21, 211301 (2013)
  [arXiv:1207.3342 [gr-qc]].
  
\bibitem{GreenSchwarzWitten}
M.~B.~Green, J.~H.~Schwarz, and E.~Witten, ``Superstring Theory'', Cambridge University Press, (1998).


\bibitem{Shannon:1948dpw}
C.~E.~Shannon,
``A mathematical theory of communication'',
Bell System Technical Journal {\bf 27}, no.3, 379-423 (1948); 
R.~V.~L.~Hartley, ``Transmission of Information", Bell System Technical Journal {\bf 7}, no.3, 535-563 (1928). 


\bibitem{Aharonov:1997md}
Y.~Aharonov, J.~Oppenheim, S.~Popescu, B.~Reznik and W.~G.~Unruh,
``Measurement of time of arrival in quantum mechanics,''
Phys. Rev. A \textbf{57}, 4130 (1998)
[arXiv:quant-ph/9709031 [quant-ph]].



 
\bibitem{Caves:1994zz}
C.~M.~Caves and P.~D.~Drummond,
``Quantum limits on bosonic communication rates,''
Rev. Mod. Phys. \textbf{66}, 481-537 (1994).

 
\bibitem{Dubovsky:2012wk}
S.~Dubovsky, R.~Flauger, and V.~Gorbenko,
``Solving the Simplest Theory of Quantum Gravity'',
JHEP \textbf{09}, 133 (2012)
[arXiv:1205.6805 [hep-th]].


  
\bibitem{Bekenstein:1981zz}
J.~D.~Bekenstein,
``Energy Cost of Information Transfer,''
Phys. Rev. Lett. \textbf{46}, 623-626 (1981); 
%
  %
  %
R.~Bousso,
``Universal Limit on Communication,''
Phys. Rev. Lett. \textbf{119}, no.14, 140501 (2017)
[arXiv:1611.05821 [hep-th]].



\bibitem{WangYao2024}
J.~Wang and S.~Yao,
``Quantum Energy Transportation versus Information Teleportation'', 
too appear, 23 May 2024.
   


  
  

  

  
\end{thebibliography}
\end{document}